          [1999/12/02 1.01 (PWD)]
\documentclass[a4paper,twocolumn]{esapub} 
\usepackage{natbib}
\usepackage{psfig}
\title{ON THE SOLAR ROTATION RATE IN THE UPPER CONVECTION ZONE}
\author{Kiran Jain}
\author{S. C. Tripathy}
\author{A. Bhatnagar}
\affil{Udaipur Solar Observatory, Physical Research Laboratory, P.O. Box No. 198, 
Udaipur 313 004 (INDIA)}

\begin{document}


\maketitle

\begin{abstract}
We present results on variation in rotation rate in the upper convection zone using 
data from GONG and MDI/SOHO  covering a period of more than four years. We find that the
 first few odd-order
splitting coefficients vary systematically with the solar cycle. 
The rotation rate near the solar surface  calculated from analytical methods 
agrees well with that of inversion techniques. 
The residual rotation rate in the outer layers seem to be correlated with the solar activity.
\end{abstract}

\section{Introduction}

With the advent of helioseismology and specially over  last few years
 remarkable progress has been made in characterising
the differential rotation of the sun. Analysis of the first 4 month's 
Global Oscillation Network Group (GONG) data
\citep{thomp96} have qualitatively confirmed  
the results of earlier analyses that the surface differential rotation
changes to rigid rotation below the outer 30\% of the solar radius.
 In addition, data from GONG network and medium-$\ell$ 
 program of Michelson Doppler Imager (MDI) on board Solar and Heliospheric Observatroy
(SOHO)  have allowed more detailed 
study of the rotational dynamics. This is achieved due to the splitting of the sun's
global oscillation frequencies by large scale flows. Details of this can be found in a recent review  by \citet{howe98}. 

	\citet{wod93} studied the time variation of equatorial rotation rate for 
different $\ell$ ranges and suggested a small variation from year to year. Inversions
for the rotation rate in solar interior have confirmed that the temporal variations
in the rotation rate penetrate to some what deeper layers \citep{hkh00,toomre00,antia00}	
Further, the changing pattern of solar rotation rate in the interior also agrees
with the torsonal oscillations observed at the solar surface \citep{howard80}. These
zonal variations of the sun's differential rotation were first observed by 
\citet{sasha97} using MDI {\it f}-mode data. \citet{toomre00} have confirmed these zonal flows using both
{\it p}- and {\it f}-mode splittings derived from MDI data. 

	In this paper, we analyse the odd order splitting coefficients  for 
detecting any periodic or systematic variation with solar cycle. 
Extending the analytical formulation of \citet{mor88} to include coefficients up to 9th order, we  infer the solar ratation rate 
as a function of depth and latitude. 
These rotation profiles are compared with those obtained from a  1.5d inversion 
technique. %
We also estimate the variation in the rotation rate from the solar minimum 
to maximum period of the current cycle. 

\section{The Data and Analysis Technique}
It is known that the solar differential rotation and other symmetry 
breaking factors like magnetic field can lift the degeneracy of the solar 
acoustic modes and split the eigen frequencies. Individual mode splittings
can  be represented by polynomial expansion 
\begin{equation}
\nu_{n,\ell,m} = \nu_{n,\ell} + \sum_{s=1}^{m} a_{s,n,\ell} P_s(m), 
\end{equation}
where $\nu_{n,\ell}$ is the mean multiplet frequency, $P_s(m)$'s are orthogonal polynomials
of degree $\ell$ and the expansion coefficients $a_{s,n,\ell}$ are known as  the splitting coefficients.
The odd-order coefficients  measure the solar 
 rotation while the even-order coefficients  probe the  symmetry about the
 equator. The nonzero values of these even coefficients reflect the pole--equator 
 asymmetries  in the solar structure \citep{dzi00}. 

We use data from both GONG \citep{hill96} and MDI \citep{schou99} to determine the rotation rate
in the outer convection zone. Each of the 45 GONG data sets covers a period of 108 days
and is centered on dates 36 days apart. The month 1 starts from May 7, 1995 and month 
46 ends on December 23,  1999. The MDI data consists of twenty 72 days time series starting from 
May 1, 1996 and ending on August 31, 2000 
 and includes the breakdown period of mid 1998.  In case of the GONG data sets, splitting
coefficients are fitted up to $a_9$ while in MDI data, the fits obtain coefficients  up to
$a_{18}$. 
 In the present work, we  further restrict the 
 frequency range to 1.5 $\le \nu \le$ 3 mHz and  $\ell \ge$ 70.
 
\section{Results and Discussions}
\subsection{Variations in the Splitting Coefficients}

 \citet{hkh99} studied the variation in GONG odd-order $a$ coefficients but could not find 
 significant correlation with solar activity measures. 
 Later the same authors \citep{hkh00} reported small 
 trends  up to $a_{15}$ coefficients. We have carried out a detailed analysis
 of these coefficients and observe that first few 
 odd-order coefficients (up to $a{_7}$) from MDI show significant correlation with activity
 measures (see Table-1). However, GONG odd-order coefficients reveal weaker correlation
  with activity. The temporal variation in odd-order coefficients  is shown in Figure~1.
  We notice that there is no clear trend in higher order coefficients while lower order coefficients 
  vary in a systematic way. Further, the $a_3$ coefficients from MDI reveal 
strong anti-correlation with activity while these from GONG do not show any
systematic trend. 
 
\begin{table}
  \begin{center}
    \caption{Correlation statistics for odd order coefficients $a_i$.}\vspace{1em}
    \renewcommand{\arraystretch}{1.2}
    \begin{tabular}[h]{lcccccc}
      \hline
      Activity & \multicolumn{3}{c}{GONG data }&\multicolumn{3}{c}{MDI data}\\
      \cline{2-3} \cline{5-7} 
      Index& $r_p$     &  $r_s$     &&& $r_p$& $r_s$      \\
      \hline
      &&&$a_1$&&& \\
      \cline{1-7}
      $R_I$ & 0.82 & 0.71&&& 0.91&0.87 \\
      KPMI  & 0.80& 0.72 &&& 0.90& 0.83  \\
      F$_{10}$& 0.83&0.73&&&0.92&0.87  \\
      \cline{1-7}
      &&&$a_3$&&& \\
      \cline{1-7}
      $R_I$ & 0.61 & 0.38&&& $-$ 0.89& $-$0.87 \\
      KPMI  & 0.58& 0.33 &&& $-$ 0.94 & $-$ 0.92  \\
      F$_{10}$& 0.62&0.40&&&$-$ 0.91&$-$ 0.89  \\
      \cline{1-7}
      &&&$a_5$&&& \\
      \cline{1-7}
      $R_I$ & 0.85 & 0.67&&& 0.89&0.76 \\
      KPMI  & 0.89& 0.74 &&& 0.92 & 0.77  \\
      F$_{10}$& 0.89&0.69&&&0.92&0.73 \\
      \cline{1-7}
      &&&$a_7$&&& \\
      \cline{1-7}
      $R_I$ & 0.92 & 0.83&&& 0.84&0.76 \\
      KPMI  & 0.94& 0.87 &&& 0.84 & 0.79 \\
      F$_{10}$& 0.94&0.83&&&0.86&0.79  \\
       \hline 
\\
      \end{tabular}
    \label{tb1}
  \end{center}
\end{table}

\subsection{Solar Rotation Rate}

The solar rotation rate using the helioseismic data are generally obtained through two different 
methods: in {\it forward approach} the frequency splittings
are computed for a chosen solar rotation model and then compared with the
observed splittings. In the {\it inverse method}  the
measured frequency splittings are used directly to produce  a single
function for the angular velocity.
In this study,  we use the analytical method of \citet{mor88} by including
higher order splitting coefficients up to $a_9$ where 
the appropriate combination of odd order splitting coefficients reflects the depth variation 
of angular velocity at chosen latitude~$\theta$  
(see Tripathy, Jain, \& Bhatnagar, 2000 for details).
\input epsf
\begin{figure}
\centering
\begin{center}
\epsfxsize=3.2in \epsfbox{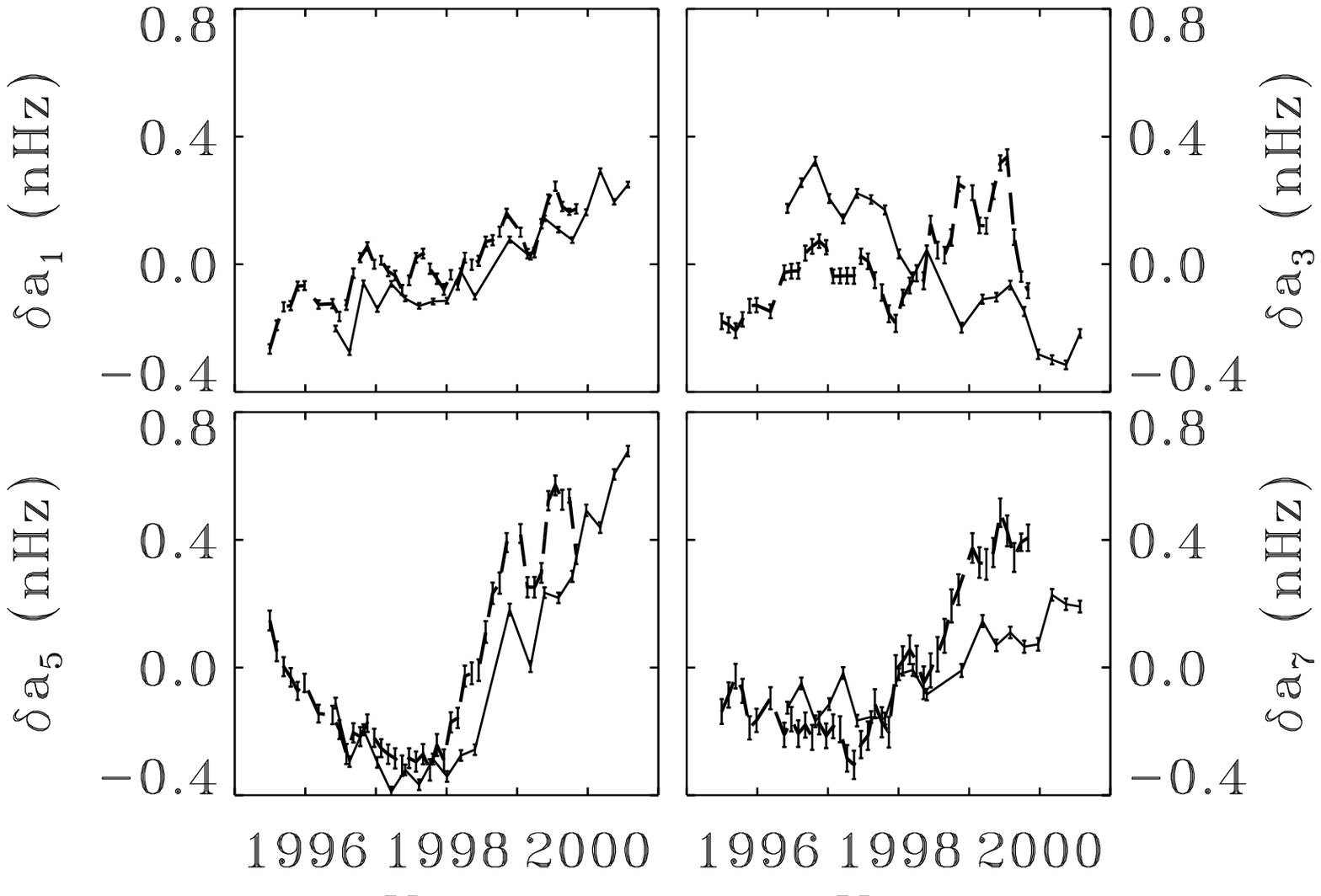}
\epsfxsize=3.2in \epsfbox{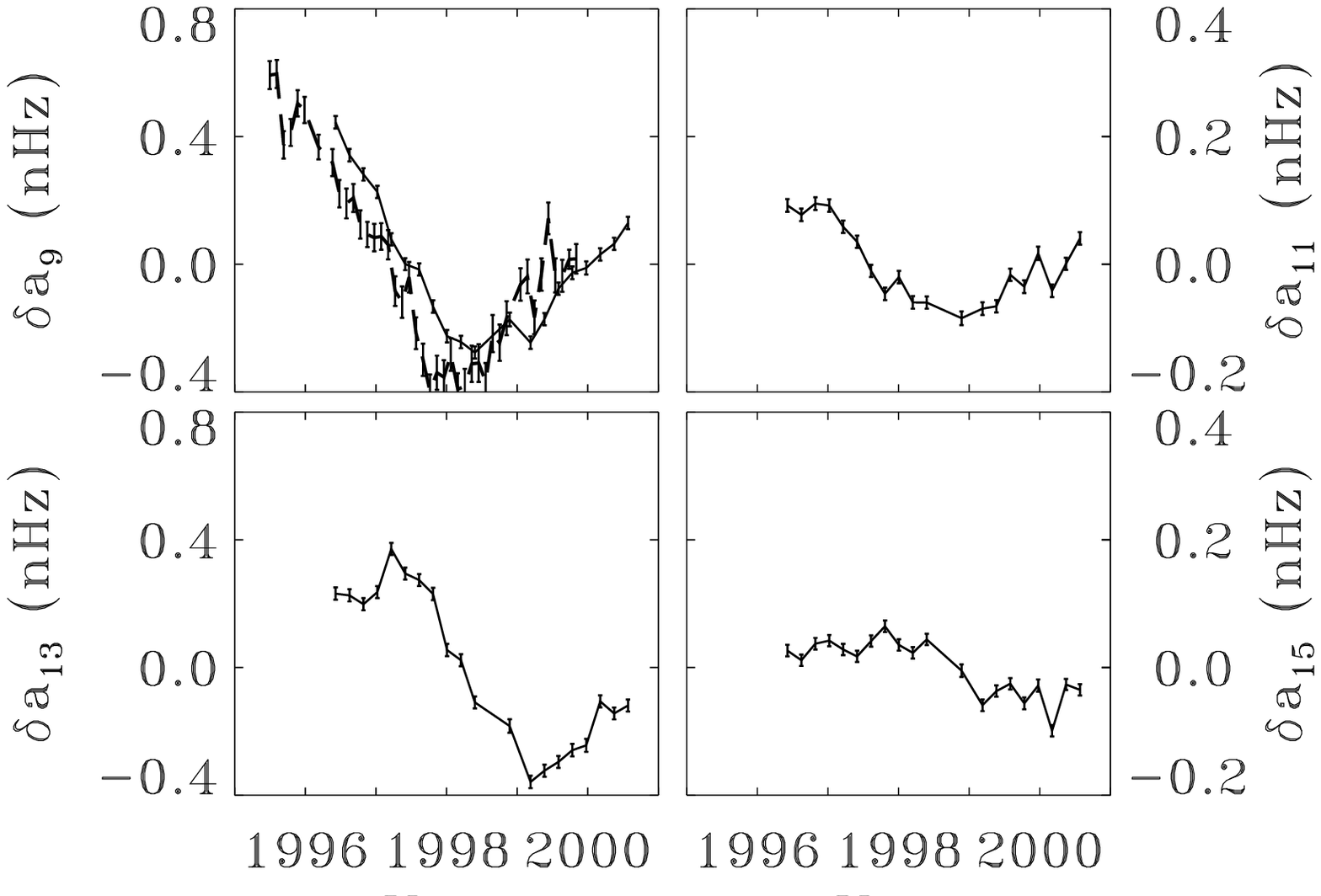}\vspace{0.5cm}
\caption{The variation of mean odd--order splitting coefficients with time. The MDI data is 
represented by solid line while the dashed line shows the GONG coefficients. 
 The 1$\sigma$ error bars  in MDI data are not clearly visible since these are very small.  
\label{fig1}}
\end{center}
\end{figure}

\begin{figure}
\centering
\begin{center}
\epsfxsize=3.2in \epsfbox{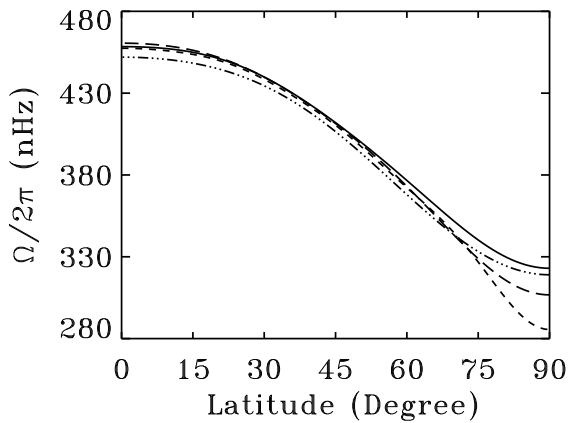}
\vspace{0.5cm}
\caption{The variation of average solar rotation rate calculated from five term fits 
in even power of cos $\theta$ are shown as a function of latitude. The line style have the following 
meaning: solid line  for \citet{snod84},  
  long dashed  line for GONG, dash-dot-dot-dot line for  MDI  
and short dashed for GONG using 1.5d RLS inversion technique.
\label{fig2}}
\end{center}
\end{figure}

It is conventional to express the observed rotation rate in terms of even powers of cos$\theta$:
\begin{equation} 
\Omega(\theta) = \sum_{i=0}^{m} A_{i} \  {\rm cos}^{2i}(\theta),
\end{equation} 
where $\theta$ is the latitude and  $m$ is restricted to 2. With the availability of
 higher order $a$-coeffcients,
we extend and retain terms up to $m$ = 4. In the asymptotic limit, we can express 
surface constants $A_i$ in terms of splitting
coefficients as: 
 \begin{eqnarray} 
 A_0&=& a_1 +a_3 + a_5 + a_7 + a_9,\\ 
 A_1 &= & -\ [5 a_3 + 14 a_5 + 27 a_7 + 44 a_9], \\
 A_2 &=& 21 a_5 + 99 a_7 + 286 a_9, \\
 A_3 &=& -\ [ {\frac{429}{5}} a_7 + 572 a_9], \\
 A_4 &=&  {\frac{2431}{7}} a_9. 
\end{eqnarray}

Various authors have calculated these surface coefficients using
different data sets and considering only three terms in the expression (2).  
\citet{brw89} found $A_0$ = 462.8 nHz, $A_1$ =
$-$~56.7 nHz and $A_2$~=~$-$~75.9 nHz for $r~\ge~$0.723~$R_\odot$ using
CaII K intensity data taken from South pole.  Based on 100-day
observations made at BBSO for $\ell$ between 10 to 60, \citet{lib89}
found the best fit with $A_0$~=~461 nHz, $A_1$ = $-$ 60.5 nHz and $A_2$~=~$-$~75.4 nHz.  We have derived these constants from the GONG data for the 
70 $\le$ $\ell$  $\le$ 150 and
obtained the best fit with $A_0$= 460.55 nHz, $A_1$ = $-$~55.76 nHz and
$A_2$ = $-$~154 nHz, $A_3$ =  220.36 nHz and $A_4$ =~$-$~164.43 nHz. 
Similarly, MDI data yields $A_0$= 458.43 nHz, $A_1$ = $-$~40.58 nHz and
$A_2$ = $-$~182.81 nHz, $A_3$ =  223.893 nHz and $A_4$ = $-$~136.09 nHz.
 The values obtained for $A_0$  are in close agreement with the earlier values.

In Figure~2, we plot the average surface rotation rate as a function
 of latitude using the derived coefficients. In the same plot, we have also shown the results from
 1.5d RLS inversion  and Doppler surface 
 measurements \citep{snod84}.
 It  is clear that the rotation rate changes significantly in mid latitude
 while the change in rotation rate near pole and  equator is
 small. We find that the rotation rate derived from the analytical approach 
 agrees  well with other results. However, the inverted rotation profile 
 departs from other rotation rates beyond the latitude of 70 degree, probably due to the 
 resolution limitation in inversion techniques. 

Figure~3 shows the residual surface rotation rate at two
 different latitudes  
obtained after subtracting the average angular velocity. 
It is evident that at higher 
latitudes, the residual rotation rate, commonly known as zonal flows, is time dependent. 
This has a magnitude of 
approximately 3 nHz at equator. The residual rotation rate at equator for two different 
depths is calculated using 1.5d RLS inversion \citep{antia00a} and is illustrated in Figure~4. 
 
It is clear that in the outer layers, the residual flows has a systematic variation
while at deeper layers, the variation does not appear to be systematic.
Recently \citet{antia00}, \citet{hkh00} using GONG data, and 
  \citet{toomre00} using  MDI data, have also reported a small but significant time 
  variation in the rotation rate. 

\section{CONCLUSION}

The first few odd-order 
coefficients are found to vary systematically with the solar cycle. We 
also detect a small but significant variation in the
rotation rate derived from the linear combination of odd order
coefficients over a period of four and half years.  The residual rotation rate 
in the outer layers seem to be correlated with the solar activity while no such
signature is found at the base of the convection zone.

\begin{figure}
\centering
\begin{center}
\epsfxsize=3.2in \epsfbox{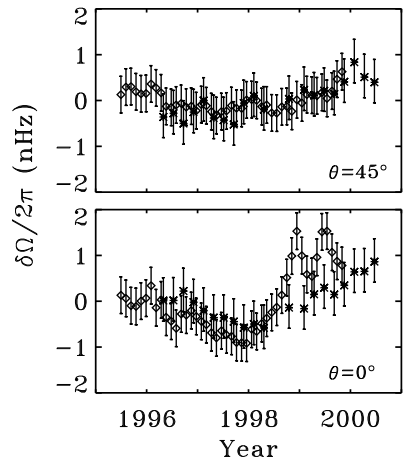}
\vspace{0.5cm}
\caption{The temporal evolution of  residual rotation rate near the surface at 
two different latitudes using both GONG(diamonds) and 
MDI data (stars).  \label{fig3}}
\end{center}
\end{figure}

\begin{figure}
\centering
\begin{center}
\epsfxsize=3.2in \epsfbox{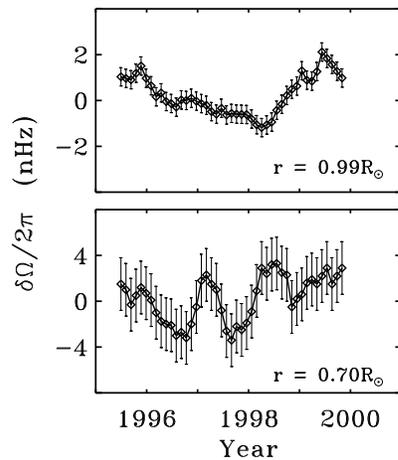}
\vspace{0.5cm}
\caption{The temporal evolution of the residual rotation rate at equator
for two different radii as inferred from GONG data using 1.5d RLS inversion technique.   \label{fig4}}
\end{center}
\end{figure}

\section*{Acknowledgments}
We thank H. M. Antia for allowing us to use the inversion code. This work utilises data obtained by the Global Oscillation Network Group project, managed by the National Solar 
Observatory, a Division of the National Optical Astronomy Observatories, which is operated by 
AURA, Inc. under cooperative agreement with the National Science Foundation. The data were 
acquired by instruments operated by Big Bear Solar Observatory, High Altitude Observatory, 
Learmonth Solar Obsrvatory, Udaipur Solar Observatory, Instituto de Astrophsico de Canaris, and 
Cerro Tololo Interamerican Observatory. This work also utilises data from the Solar Oscillations Investigation / Michelson Doppler Imager
 on the Solar and Heliospheric Observatory and we thank J. Schou for providing us 
the data sets. 
 SOHO is a mission of international cooperation
 between ESA and NASA. NSO/Kitt Peak magnetic, and Helium measurements used 
here are produced cooperatively by NSF/NOAO; NASA/GSFC and NOAA/SEL.  This work is 
partially supported under the CSIR Emeritus Scientist Scheme and Indo--US collaborative 
programme--NSF Grant INT--9710279. 

\end{document}